\documentclass[12pt]{article}
\usepackage{a4wide}
\usepackage{graphics}
\usepackage{graphicx}
\usepackage{amsmath}

\newcommand{\bbbone}{\hbox{\rm 1\kern-3pt l}}
\newcommand{\Jp}{J/\psi}

\newcommand{\GeV}{\mbox{\,GeV}}
\newcommand{\TeV}{\mbox{\,TeV}}
\newcommand{\mb}{\mbox{\,mb}}
\newcommand{\mub}{\mbox{\,$\mu$b}}
\newcommand{\nb}{\mbox{\,nb}}
\newcommand{\pb}{\mbox{\,pb}}

\newcommand{\seff}{{\sigma_{\rm eff}}}
\newcommand{\Rd}{R_{\Delta}}

\begin{document}

\thispagestyle{empty} 

\begin{center}
{\Large\bf Double $J/\psi$ production as a test of 
parton momentum correlations in double parton scattering}\\[1cm]
{\large Sergey Koshkarev and Stefan Groote}\\[0.3cm]
Institute of Physics, University of Tartu, 51010 Tartu, Estonia\\[3pt]
\end{center}

\begin{abstract}
Using the GS09 model we predict the possible impact of the parton momentum
correlation on the $J/\psi$-pair production at the Spin Physics Detector at the
Nuclotron-based Ion Collider Facility. The double $J/\psi$ production 
and the effective cross sections are calculated.
\end{abstract}

\section{Motivation}

With the advent of high luminosity accelerators, multi parton scattering (MPS)
physics at high-energy hadron colliders like the Tevatron at Fermilab and the
Large Hadron Collider (LHC) at CERN, in particular those observing double
parton scattering (DPS), has become one of the hottest topics of the modern
experimental particle physics~\cite{Akesson:1986iv,Alitti:1991rd,Abe:1993rv,%
Abe:1997xk,Abazov:2009gc,Aaij:2012dz,Aad:2013bjm,Chatrchyan:2013xxa,%
Abazov:2014fha,Abazov:2014qba,Aad:2014kba,Aaij:2015wpa,Abazov:2015nnn,%
Abazov:2015fbl,Aaboud:2016dea,Khachatryan:2016ydm,An:2017kyn,Belyaev:2017sws,%
Aaij:2016bqq,Sirunyan:2017hlu,Aaboud:2018tiq,CMS:2019jcb,LHCb:2020jse,%
CMS:2021wfx,CMS:2021lxi,CMS:2022pio,Leontsinis:2022cyi}. DPS results are
usually interpreted in terms of the so-called ``pocket
formula''~\cite{dEnterria:2017yhd}
\begin{equation}
\label{eq:pocket}
\sigma^{AB}_{\rm DPS}=\frac{m}{2}
  \frac{\sigma^A_{\rm SPS}\sigma^B_{\rm SPS}}{\seff},
\end{equation}
where the combinatorial factor is $m = 1$ for identical final states $A=B$ and
$m = 2$ for $A\neq B$. This formula is derived under the assumption of
independent parton scatterings, with
\begin{equation}
\label{eq:s_eff}
\seff = \bigg[\int d^2b~T^2({\bf b})\bigg]^{-1},\qquad
  \int d^2b~T({\bf b})=1,
\end{equation}
where $T({\bf b})$ is the overlap function that characterizes the transverse
area occupied by the interacting partons in the impact parameter space
${\bf b}$~\cite{dEnterria:2017yhd}. Such a definition leads to the assumption
that the value of $\seff$ should be universal, independent of the final states
or the phase space. On the other hand, the D\O\ collaboration provided two
different measurements of DPS for $\gamma + 3$~jet events, namely
$\langle\seff\rangle = 16\pm0.3\pm2.3\mb$ in 2010~\cite{Abazov:2009gc} and
$\sigma_{\rm eff}^{\rm incl} = 12.7\pm0.2\pm1.3\mb$ in 2014~\cite{Abazov:2014fha}.
It is easy to see that not only the central values are different, but the
errors bars do not overlap neither. In contrast to many other results, the
only difference between these two measurements is the kinematic region (cf.\
Tab.~\ref{tab:dps_overview_Tevatron}). This definitely contradicts the
``universality'' assumption for $\seff$.

In Refs.~\cite{Gaunt:2009re,Snigirev:2010tk,Rinaldi:2014ddl} the importance of
the effect of the evolution of the double parton distribution function (dPDF)
is emphasized (for a detailed definition of the dPDF cf.\ e.g.\
Ref.~\cite{Diehl:2011yj}). Most of the phenomenological analysis of a possible
impact of the dPDF effect are focused on the LHC energies (cf.\
Refs.~\cite{Ceccopieri:2017oqe,Gaunt:2010pi,Kom:2011bd,Kom:2011nu,%
Borschensky:2016nkv} and references therein). Indeed, standard wisdom tells us
that DPS should be more preferable at high energies. On the other hand,
possible dPDF effect is expected at higher values of the Bjorken-$x$ or with a
significant gap between the $x$ values. However, as it was discussed in 
Ref.~\cite{Ceccopieri:2017oqe} at $\sqrt{s} = 13\TeV$, even for double W-boson
production the typical value for the Bjorken-$x$ is
$\langle x \rangle \sim 0.01$. In contrast to that, preferable conditions can
be easily achieved at lower energy experiments, where the DPS contribution
could be relatively small but still expected to be far from
zero~\cite{Lansberg:2015lva,Koshkarev:2019crs}.

The Spin Physics Detector at the Nuclotron-based Ion Collider fAcility (NICA)
collider (JINR, Dubna) is a universal facility to investigate the spin
structure of the proton and deuteron and the other spin-related phenomena with
polarized proton and deuteron beams at a collision energy up to $27\GeV$ and a
luminosity up to $10^{32}$ ${\rm cm^{-2} ~ s^{-1}}$~\cite{SPDproto:2021hnm}.
It is easy to see that the production threshold of the pair of $\Jp$ is
already more than $20\%$ of the NICA energies, leading to the typical value
for the Bjorken-$x$ of $\langle x \rangle > 0.1$. The relatively small cross
section can be compensated by the high luminosity.

In this paper we phenomenologically investigate the possible impact of the
evolution of the dPDF on the production cross section of double $\Jp$
events and the measurements of $\seff$ at NICA energies.

\begin{table}[ht]
\caption{\label{tab:dps_overview_Tevatron} The effective cross section measured 
in $\gamma + 3$~jets process in $p \bar{p}$ collisions at $\sqrt{s}=1.96\TeV$
measured by the D\O\ experiment at the Tevatron.}\vspace{7pt}
\begin{center}\begin{tabular}{ |c|c|c|c|c|c| } 
\hline
Year&Final State&$p_T^{\rm min}(\GeV/c)$&$\eta(y)$ range&$\seff$ (mb)\\
\hline
\begin{tabular}{@{}l@{}}2010\\\cite{Abazov:2009gc}\end{tabular}&
$\gamma + 3$~jets&
\begin{tabular}{@{}c@{}}$60<p_T^{\gamma}<80$\\
  $p_T^{\rm jet\,1} > 25$\\$p_T^{\rm jet\,2,3} > 15$\end{tabular}&
\begin{tabular}{@{}c@{}}$|y^{\gamma}|<1.0$\\$1.5<|y^{\gamma}|<2.5$\\
  $|y^{\rm jet}|<3.0$\end{tabular}&$\langle \seff \rangle = 16\pm0.3\pm2.3$\\
\hline
\begin{tabular}{@{}l@{}}2014\\\cite{Abazov:2014fha}\end{tabular}&
\begin{tabular}{@{}c@{}}$\gamma + 3$~jets\\$\gamma + c/b + 2$~jets\end{tabular}&
\begin{tabular}{@{}c@{}}$p_T^{\gamma}>26$\\$p_T^{\rm jet\,1} > 15$\\
  $15<p_T^{\rm jet\,2,3}<35$\end{tabular}&
\begin{tabular}{@{}c@{}}$|\eta^{\gamma}|<1.0$\\$1.5<|\eta^{\gamma}|<2.5$\\
  $|\eta^{\rm jet}|<2.5$\end{tabular}&
\begin{tabular}{@{}c@{}}$\seff^{\rm incl} = 12.7\pm0.2\pm1.3$\\
  $\seff^{\rm HF} = 14.6\pm0.6\pm3.2$\end{tabular}\\
\hline
\end{tabular}\end{center}
\end{table}

\section{Phenomenological basement}

In single parton scattering, the evolution of the sPDF is described by the
DGLAP QCD evolution equations, formerly known as the Altarelli--Parisi
equations. As shown in Ref.~\cite{Ellis:1991qj}, these renormalisation group
equations can be solved by imposing two sum rules, namely the momentum and the
number sum rule. While for single parton scattering the sums over momenta
fractions and the number of valence partons stay constant, the sum rules for
double parton scattering are more involved and, as it turns out, more
restrictive.

Corresponding to the (s)DGLAP for single parton scattering, Gaunt and Stirling
postulate a dDGLAP which is based on a couple of principles explained in
detail in Ref.~\cite{Gaunt:2009re}. As for the sDGLAP, an increasing scale
$t=\ln(Q^2)$ allows for the splitting of partons (by, e.g., the emission of a
gluon), described by the splitting functions $P_{i\to j}(x)$ which denote a
parton $i$ splitting to $j$ with a momentum fraction $x$. For double parton
scattering, the splitting functions $P_{i\to jk}(x)$ can be understood as the
parton $i$ splitting to a parton $j$ with momentum fraction $x$, and a parton
$k$ with momentum fraction $1-x$. This latter is certainly true in leading
order (LO) scattering but has to be modified in next-to-leading order (NLO).
For this reason, in Ref.~\cite{Gaunt:2009re} Gaunt and Stirling deal with the
LO scattering only, postponing the NLO analysis to a future publication.
In LO, the sum rules to be postulated have to take into account two very basic
correlations, namely
\begin{enumerate}
\item having found a quark with a given flavour, the probability is smaller
  to find a second quark with the same flavour, and
\item having found a parton with momentum fraction $x$, the probability is
  smaller to find a second parton with momentum faction $1-x$.
\end{enumerate}
Based on three splitting diagrams for increasing and two diagrams for
decreasing population of partons, the dDGLAP equation is justified in detail
(cf.\ Eq.~(2.1) and Fig.~3 in Ref.~\cite{Gaunt:2009re}). The new sum rules are
postulated, and a procedure is developed to solve the dDGLAP on a grid of
momentum fractions and scale, known as the implementation of the GS09 model.

If the final states $A$ and $B$ are produced in a DPS process independently,
Eq.~(\ref{eq:pocket}) can be phenomenologically cast into the form
\begin{eqnarray}
\sigma^{AB}_{\rm DPS} & = & \frac {m}{2} \frac 1 \seff \int dx_1 \ldots dx_4
  \ ~\ f(x_1,Q_A) f(x_2,Q_A) \hat{\sigma}_A(x_1, x_2) \times \nonumber \\ &&
  f(x_3,Q_B) f(x_4,Q_B) \hat{\sigma}_B(x_3, x_4) 
  \theta (1 - x_1 - x_3) \theta (1 - x_2 - x_4),
\end{eqnarray}
where $f(x,Q)$ denotes the parton distribution function, $\hat{\sigma}$ is
the cross section at parton level, and $\theta$ is the Heaviside step
function.\footnote{We use odd indexes for the ``first'' proton and even
indexes for the ``second'' proton.}

As we already mentioned, some models predict the correlation between partons
of a same hadron~\cite{Gaunt:2009re,Rinaldi:2014ddl}. In this case, 
Eq.~(\ref{eq:pocket}) should be written in the form
\begin{equation}
\label{eq:dps_orig}
\sigma^{AB}_{\rm DPS} = \frac {m}{2} \frac {1}{\seff} \int dx_1 \ldots dx_4
  ~ D(x_1, x_3, Q_A, Q_B) D(x_2, x_4, Q_A, Q_B)
  \hat{\sigma}_A(x_1, x_2) \hat{\sigma}_B(x_3, x_4),
\end{equation}
where e.g.\ the dPDF
$D(x_1, x_3, Q_A, Q_B) \neq  f(x_1,Q_A) f(x_3,Q_B) \theta (1 - x_1 - x_3)$
is not the product of the two sPDFs. For the possible impact of the dPDFs we
can write the ratio
\begin{equation}
\label{eq:Rd}
\Rd (x_1, x_2, x_3, x_4, Q_A, Q_B)
  = \frac{D(x_1, x_3, Q_A, Q_B) D(x_2, x_4, Q_A, Q_B)}
  {f(x_1,Q_A) f(x_2,Q_A) f(x_3,Q_B) f(x_4,Q_B)}.
\end{equation}

\section{$\Jp$-pair production from DPS with the ``pocket formula''}

As a first step, it is interesting to estimate the DPS effect with the
``pocket formula'' approximation~(\ref{eq:pocket}). In order to provide such
an estimate, we are going to use experimentally measured $\Jp$ production
cross sections at similar energies. Using the CERN proton beam at $400\GeV$/c
to produce charm particles with incident on different nuclear targets, the NA3
experiment provided data on the production of $\Jp$ pairs on a platinum target
with the production cross sections of $27 \pm 10 \pb$ per
nucleon~\cite{NA3:1985rmd}, and the NA50 experiment measured single $\Jp$ 
production $Br(\Jp \to \mu^+ \mu^- ) \times \sigma(\Jp)$ on Be, Al, Cu, Ag, W,
and Pb targets in the range between $4.717 \pm 0.026 \nb$ and
$3.715 \pm 0.016 \nb$ per nucleon~\cite{NA50:2006rdp}.

Combining the mean values of the single $\Jp$ cross section for proton-W and
proton-Pb collisions\footnote{We assume that all single $\Jp$ produced via the
process $gg \to \Jp$, i.e., any other possible mechanism is not seen at any
evidence level with provided accuracy. This assumption is strongly supported
by the analysis in Ref.~\cite{NA50:2006rdp}},
$\sigma(\Jp) \approx 12.5\mub$, and the effective cross
section $\langle \seff \rangle = 4.6 \mb$ for double $\Jp$ production (cf.\
Tab.~\ref{tab:cs_eff}), we can calculate the contribution of DPS. However,
$\sigma(\Jp)$ as measured by the NA50 collaboration includes $\Jp$ from the
feed-down effect, i.e., the final $\Jp$ state could be a result not only of
prompt $\Jp$ production but also of the decay of higher mass states,
$\psi' \to \Jp + X$ and $\chi_c \to \Jp + \gamma$. To consider this effect, we
cast formula~(\ref{eq:pocket}) into the form
\begin{equation}
\sigma_{DPS}(\Jp\Jp) = \frac{\sigma(\Jp)^2}{\seff}
\bigg(\frac{r^2_{\Jp} + r^2_{\psi'} + r^2_{\chi_c}}{2}
+ 2 \cdot(r_{\Jp}r_{\psi'} +  r_{\Jp}r_{\chi_c} + r_{\psi'}r_{\chi_c})\bigg)
\end{equation}
(cf.\ also the discussion in Ref.~\cite{Lansberg:2014swa}), where
$r_{\Jp} = 0.62\pm0.04$, $r_{\psi'} = 0.08\pm0.02$, and
$r_{\chi_c} = 0.30\pm0.08$ are feed-down fractions~\cite{Digal:2001ue}.
Performing this calculation, we obtain $\sigma_{DPS}(\Jp\Jp) = 2.6\pb$. 
Utilizing the double $\Jp$ production cross section measured by the NA3
experiment, the fraction of DPS events can be estimate to be
$f_{\rm DP} \approx 0.096$.
\begin{table}[ht]
\caption{\label{tab:cs_eff} The effective cross section values measured 
in double and triple $\Jp$ production at D\O, ATLAS, and CMS experiments.}
\vspace{7pt}
\begin{center}\begin{tabular}{ |c|c|c|c| } 
\hline
Experiment&$\sqrt{s}$, $\TeV$&Colliding Mode&$\seff$, $\mb$\\
\hline
D\O~\cite{Abazov:2014qba}&1.96&$p\bar{p}$
  &$4.8\pm0.5\text{(stat)}\pm2.5\text{(syst)}$\\\hline
ATLAS~\cite{ATLAS:2016ydt}&8&$pp$&$6.3\pm1.6\text{(stat)}\pm1.0\text{(syst)}$\\
\hline
CMS~\cite{Leontsinis:2022cyi}&13&$pp$
  &$2.7^{+1.4}_{-1.0}\text{(exp)}^{+1.5}_{-1.0}\text{(th)}$\\
\hline
\end{tabular}\end{center}
\end{table}
\begin{center}\begin{figure}[ht]
\centerline{\includegraphics[scale=0.6]{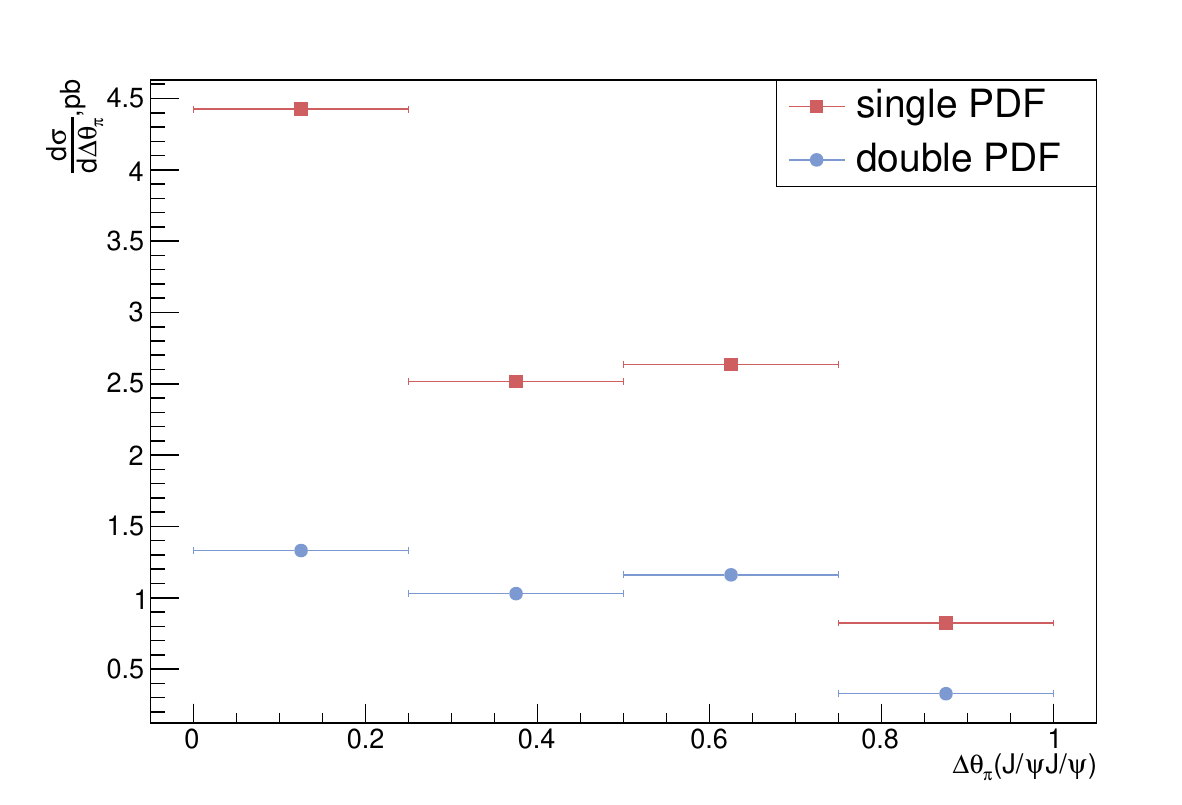}}
\caption{\label{fig:delta_Theta} Differential cross section distributions 
for double $\Jp$ production, as predicted by single PDF (squared points)
and double PDF (circular points).}
\end{figure}\end{center}

\section{Possible effects of double PDF within the GS09 model}

In order to distinguish between single and double PDF predictions, we use a
Pythia~8 Monte-Carlo simulation~\cite{Sjostrand:2007gs}, where the sPDFs are
calculated with MSTW2008LO~\cite{Martin:2009iq} and the dPDFs are calculated
in the GS09 model employing the ratio~(\ref{eq:Rd}), where $\Rd$ was
calculated for every single event.

The effect of double PDF for double $\Jp$ production cross section at NICA,
differential in the normalized angular difference for the polar angle, 
$\Delta\theta_\pi = (\theta(\Jp_1) - \theta(\Jp_2))/\pi$, is presented in
Fig.~\ref{fig:delta_Theta}. In case of the single PDF prediction, the first
bin $\Delta\theta_\pi < 0.25$ contains more than $40\%$ of the statistics.
In contrast to that, double PDF predicts almost equal statistics for all bins,
except for the bin $\Delta\theta_\pi > 0.75$, where almost no events are found. 

It is also interesting to calculate $\seff$ in the style of a ``pocket
formula'' analysis. Our calculation gives $\langle \Rd \rangle \approx 0.37$
that leads us to the effect of the dPDFs, namely the amplification of
$\seff = \langle \seff \rangle  / \langle \Rd \rangle \approx 12.4 \mb$. This
value is much higher than the value previously measured by D\O, ATLAS and CMS
at low Bjorken-$x$. Obviously, our result also reduces the DPS fraction to
$f_{\rm DP} \approx 0.036$. We would like to remind the readers that such a
DPS fraction is comparable with respective DPS fractions for the recent LHC 
analysis (cf. Tab.~\ref{tab:fDP})
\begin{table}[ht]
\caption{\label{tab:fDP}Estimated values for $f_{\rm DP}$}\vspace{12pt}
\begin{tabular}{|c|c|c|c|c|}\hline
Ref.&events&$p_T$ etc.&$y(\eta)$&$f_{\rm DP}$\\\hline\hline
\cite{Aad:2013bjm}
&$pp\to W+2$\ jet
&$p_T>20\GeV/c$&$|y|<2.8$&$0.08\pm 0.01\pm 0.02$\\
&&$\sqrt s=7\TeV$&&\\\hline
\cite{Chatrchyan:2013xxa}
&$pp\to W+2$\ jet
&$p_T>20\GeV/c$&$|\eta|<2.0$&$0.055\pm 0.002\pm 0.014$\\
&&$\sqrt s=7\TeV$&&\\\hline
\cite{Aaboud:2016dea}
&$pp\to 4$\ jet
&$p_T\ge 20\GeV/c$&$|\eta|<4.4$&$0.092%
  \genfrac{}{}{0pt}1{+0.005}{-0.011}%
  \genfrac{}{}{0pt}1{+0.033}{-0.037}$\\
&&$\sqrt s=7\TeV$&&\\\hline
\cite{ATLAS:2016ydt}
&$pp\to J/\psi+J/\psi$
&$p_T>8.5\GeV/c$&$|y|<2.1$&$0.092\pm 0.021\pm 0.005$\\
&&$\sqrt s=8\TeV$&&\\\hline
\cite{Aaboud:2018tiq}
&$pp\to 4$\ lepton
&$80<m_{4\ell}<1000\GeV/c^2$&&$f_{\rm DP}\le 0.042$\\
&&$\sqrt s=8\TeV$&&\\\hline
\end{tabular}
\end{table}
%
  
\section{Discussion and Summary}

Current theoretical and experimental researches on parton momentum
correlations mainly focus on LHC energies. In this paper we show that lower
energy colliders like NICA also could be helpful. We propose to measure the
double $\Jp$ production cross section at NICA energies, differential in the
normalized angular difference for the polar angle,
$d \sigma_{DPS}(\Jp\Jp) / d \Delta\theta_\pi$. To avoid a precision fit of
data that could turn out to be quite speculative at the relatively low
statistics (cf.\ the discussion below) we propose to investigate the ratio
\begin{equation}
N(\Delta\theta_\pi < 0.25) / N(\Delta\theta_\pi > 0.25),
\end{equation}
where $N(\cdots)$ is the number of events in the respective kinematical
region. As we can learn from Fig.~\ref{fig:delta_Theta}, for the ``pocket
formula'' this ration is equal to $\sim 2/5$ and for the GS09 model equal to
$\sim 1/3$.

\begin{center}\begin{figure}[ht]
\centerline{\includegraphics[scale=0.6]{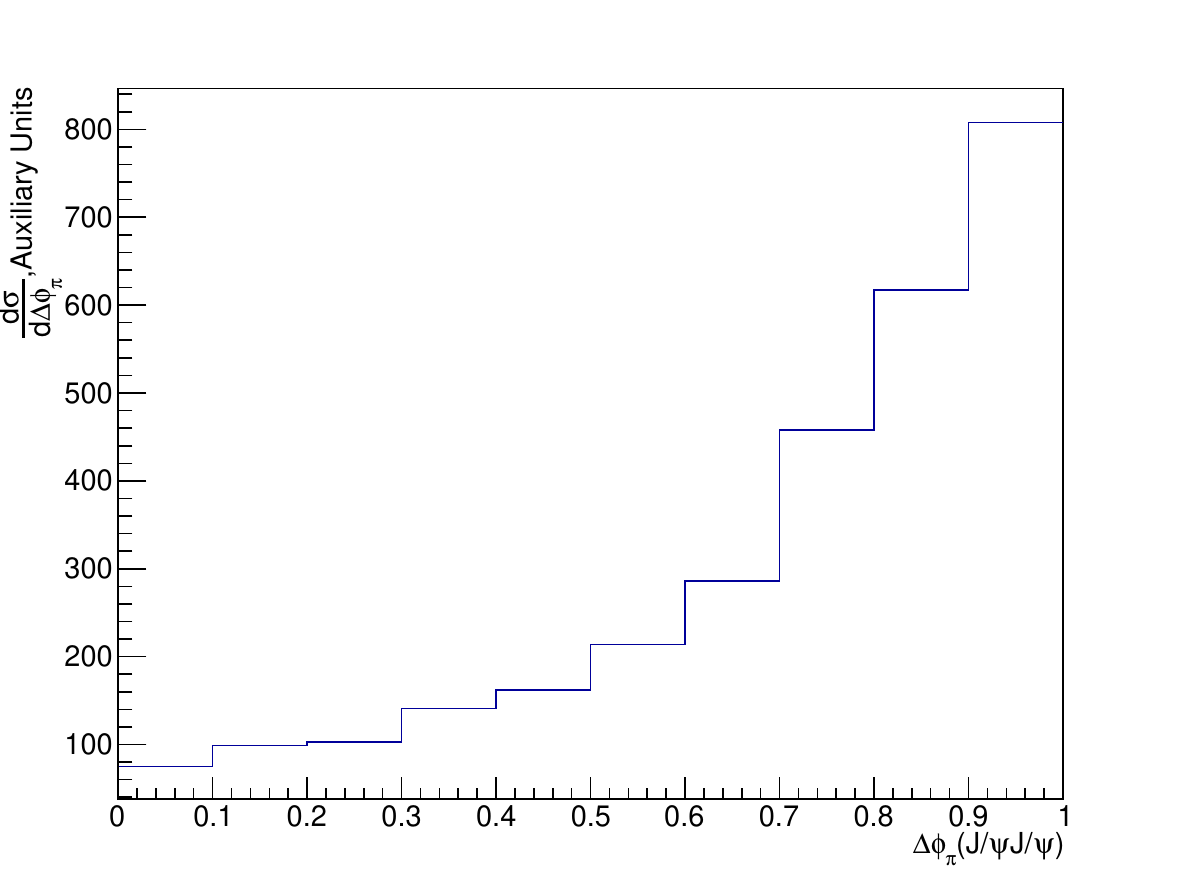}}
\caption{\label{fig:delta_phi} Differential cross section distribution for
double $\Jp$ production with LO SPS with the effect of $k_T$
smearing~\cite{Sridhar:1998rt}. We use Pythia~8 (LO calculations) and
$\langle k_T \rangle = 0.7\GeV$~\cite{NA3:1985rmd}.}
\end{figure}\end{center}

In order to distinguish DPS from SPS, we proposed to use the (normalized)
angular difference for the azimuthal angle of radiation of the $\Jp$ pair,
given by $\Delta\phi_\pi = (\phi(\Jp_1) - \phi(\Jp_2))/\pi$. Having taken into
account the fact that $\Delta \phi_\pi$ has a peak near 1 for SPS 
(cf.\ Fig.~\ref{fig:delta_phi}) but a flat shape for DPS, we were able 
to exclude the region $\Delta \phi_\pi \sim 1$ in order to maximize the
DPS/SPS ratio.

Last but not at least, it is interesting to estimate the total number of
events. According to estimates of the SPD collaboration, up to $12\times 10^6$ 
single $\Jp$ events are expected per year~\cite{Arbuzov:2020cqg}. Combining
double and single $\Jp$ production cross sections measured by the NA3 and NA50
experiments~\cite{NA3:1985rmd,NA50:2006rdp}, we can calculate the ratio
\begin{equation}
\frac{\sigma(\Jp\Jp)}{\sigma(\Jp)} > 10^{-4}.
\end{equation}
Multiplying this ratio with $f_{\rm DP}$, we can estimate the number of DPS 
$\Jp$-pairs per year as $>115$ and $>43$ in case of the ``pocket formula'' 
and GS09, respectively.

\subsection*{Acknowledgements}

The authors would like to acknowledge J.R.~Gaunt for his help in understanding
how to run the GS09 model code and providing code examples. We also would like
to thank D.~Bandurin and G.~Golovanov for very useful discussions about DPS,
and A.~Gridin and A.~Guskov for their comments about the Spin Physics Detector
and NICA. The research was supported by the European Regional Development Fund
under Grant No.~TK133.

\end{document}